\documentclass{article}

\usepackage{amsmath,amssymb}
\usepackage{epsfig}

\newcommand{\bu}{$\bullet$\ }
\newcommand{\be}{\begin{equation}}
\newcommand{\ee}{\end{equation}}
\newcommand{\bes}{\begin{subequations}}
\newcommand{\ees}{\end{subequations}}
\newcommand{\bea}{\begin{eqnarray}}
\newcommand{\eea}{\end{eqnarray}}
\newcommand{\bear}{\begin{equation}\begin{array}}
\newcommand{\eear}[1]{\end{array}\label{#1}\end{equation}}
\usepackage{wrapfig}

\def\lb{\linebreak[4]}
\def\ba{$$\begin{array}}
 \def\ea{\end{array}$$}

\newcommand{\fr}[2]{\dfrac{{ #1}}{{ #2}}}

\newcommand{\la}{\langle}
\newcommand{\ra}{\rangle}
\newcommand{\fn}[1]{\footnote{{#1}}}

\renewcommand{\le}{\leqslant}

\def\vep{{\varepsilon}}
\newcommand{\epe}{\mbox{$e^+e^-\,$}}
\newcommand{\ggam}{\mbox{$\gamma\gamma\,$}}
\newcommand{\egam}{\mbox{$e\gamma \,$}}

\def\cl{\centerline}

{\end{list}}
\newcounter{enumct}

\newsavebox{\fmbox}
\newenvironment{fmpage}[1]
{\begin{lrbox}{\fmbox}\begin{minipage}{#1}}
{\end{minipage}\end{lrbox}\fbox{\usebox{\fmbox}}}
\begin{document}
\date{}
%\today
\title{Two photon physics. Personal recollection }

\author{Ilya F. Ginzburg $^{a,b}$}

\maketitle
\noindent{$^{a}$ Sobolev Inst. of Mathematics, av. Koptyug, 4, Novosibirsk, 630090, Russia}\\
 {$^{b}$ Novosibirsk State University,  Pirogova str., 2, Novosibirsk, 630090, Russia}

\noindent{\it e-mail: Ginzburg@math.nsc.ru}
\begin{center}
{ \it Talk given at Int. Conference PHOTON 2015
Novosibirsk, 16/06/2015}
\end{center}
%%%%%%%%%%%%%%%%%%%%%%%%%%%%%%%%%%%%%%%%%%%%%%%%%%%%%%%%%%%%%%%%%%%%%
\begin{abstract}

The term  {\bf two--photon processes} is used  for the
reactions  in which some system of particles is produced in
collision of two photons, either real or virtual. In the
study of these processes our main goal was to suggest approach, allowing to extract from the data information on proper two--photon process separating it from mechanism which responsible for the production of photons.

Here I present my  view for history of two--photon physics.
I don't try to give complete review, concentrating mainly on works of our team  (which cover essential part of the topic) and some colleagues. My citation is strongly incomplete.
I cite here only  papers which were essential in our understanding of the problems. The choice  of presented details  is the result of my discussions with Gleb Kotkin and Valery Serbo.
\\
 1. Prehistory.\\
 2. Two photon processes at $e^+e^-$ colliders.\\
 3. Photon colliders. \\
 4. Notes on physical program.

\end{abstract}

%%%%%%%%%%%%%%%%%%%%%%%%
\section{Prehistory}
%%%%%%%%%%%%%%%%%%%%%%%%%%

\bu {\bf 30-th-60-th. High order processes of QED.}

\begin{wrapfigure}[15]{l}[0pt]{0.36\textwidth}%\vspace{-1mm}
\begin{fmpage}{0.33\textwidth}
\includegraphics[width=0.95\textwidth,
  height=3.8cm]{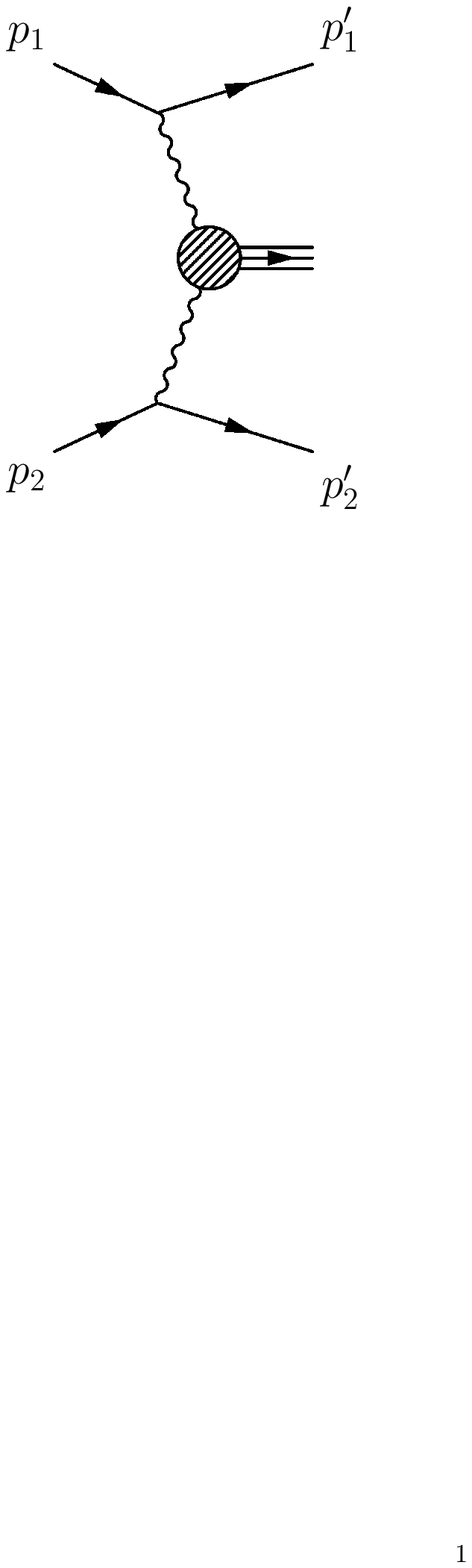}
    \caption{\it Particle production in collision of two
fast particles}
\end{fmpage}
\label{figLL}
\end{wrapfigure}

The processes which called now as two-photon ones were
discussed  after discovery of positron by Anderson (1932), when a necessity is appeared to find out the process in which
positrons are generated. In 1934 studying \epe\, pair
production in collision of ultrarelativistic charged
particles $A_1$ and $A_2$ Landau and Lifshitz \cite{LL} have ascertained
that the two photon channel of Fig.~\ref{figLL} is
dominated in this reaction. They calculated the cross section of the process\lb $A_1A_2\to
A_1 A_2\; e^+e^-$ in the leading logarithmic approximation.
Almost simultaneously Bethe and Heitler \cite{BH}
considered \epe\, pair production by photon in the field of
a nuclei, $\gamma A\to \epe A$. These processes contain subprocess $\ggam\to\epe$,
like Fig.~1.

The leading log result of \cite{LL} was improved by Racah
\cite{Racah} who have calculated the corresponding cross
section with an accuracy   $\sim (M/E)^2$ where $E$ and
$M$ are energy and mass of incident nuclei. The process
$\gamma A\to \epe A$ was included in the theory of wide
atmospheric showers in cosmic rays \cite{LRum} and in the
description of the energy losses of fast muons in matter
\cite{enloss}.

The  hadron production by two photons was considered for the
first time by Primakoff \cite{Prim} suggested in 1951 to
measure the $\pi^0$ life--time in the reaction $\gamma A\to
\pi^0 A$. The new interest to such processes was appeared when
the construction of \epe\  colliders become close to a
reality. In 1960 Low \cite{Low} pointed out
that the $\pi^0$ life--time can be measured also in the
$\epe\to \epe\pi^0$ process. Simultaneously the two--photon reaction
$\epe\to\epe\pi^+\pi^-$ (for point--like pions) was
considered \cite{CalZem}. However, the calculated rates
seemed unmeasurable small and no further work was
done  at that time.

In 1969--1970 new generation of papers appeared with the
goal to cover possible set of final states of \epe\
colliders as complete as possible. Authors considered \epe collisions with final states $\epe \pi^0$, $ \epe \eta$ and $ \epe \pi^+\pi^-$, $\epe K^+K^-$ (in the latter two cases for the point-like
pions and kaons) \cite{deCel}. Some of these processes for point-like hadrons just as $\epe\to\epe\mu^+\mu^-$
 were considered in more detail by
Paris \cite{Kess} and Novosibirsk BINP \cite{Bai} groups.
These papers did not provoke high interest in particle
physics community since they were in line with numerous
calculations of various processes at \epe\  colliders, having
small cross section (for the contemporary machines). They don't try to suggest method of extraction of information about $\ggam\to\pi\pi$, $\ggam\to KK$ subprocesses.\\

\bu {\bf End of 60-th. Popular problems.}

In the 60-th the study of different processes of hadron collisions was of main interest for community. In addition to the collisions
initiated by proton and deuton beams, the processes, initiated by pion beams, kaon beams, antiproton beams, hyperon beams (experiment and theory) were of great interest for community providing new types of final states and new field for the Regge theory developed at that time. In this respect the study of deep inelastic $ep$ scattering was a hot point in particle physics provided new type of collided hadron (photon) with variable mass and helicity.

One more popular field of studies was the coupled channel problem in the low energy scattering -- description of $\pi\pi\to\pi\pi$ and $\pi\pi\to KK$ scattering.

\section{Two photon processes at \epe \ colliders}

%%%%%%%%%%%%%%%%%
\bu {\bf Novosibirsk. 1969-1970.}
%%%%%%%%%%%%%%%%%%%%

Once in the winter 1969-1970 my PhD student Victor Budnev was informed me about observation in  Novosibirsk BINP the process $\epe\to\epe\epe$ in the group including my former student Vladimir Balakin \cite{Bal}, \cite{KievRoch}. Relatively high cross section of this 4-th order process of QED was explained by small virtuality of photons coupled  with the initial and scattered electrons. I immediately understood that similar mechanism is suitable also for the  production of hadron systems, not only for $\epe$. During few months I reported in different groups in the Moscow institutes and in the JINR about new opportunity found by experimentalists of BINP. My first proposal was to study process $\ggam\to\pi\pi$ using the methods developed for the $\pi\pi\to KK$. I  had not received a response for these proposals. And once  someone told me -- "you know, you find a new opportunity".

I understood that high energy \epe \ colliders really provide us  by opportunity to study new type of processes, yet unknown for community --  the production of particles in collisions of two photons  (I had in mind mainly production of hadrons).
The study of such process continues  investigations of deep inelastic $ep$ scattering to the  new region of parameters and final states with  two new variable parameters -- virtualities of each photon.

I invite for writing paper   V. Budnev and  V. Balakin. We tried to present  the paper which contained  the description of  processes as well as the method of extracting  the information from the data.

Fortunately, I had no experience in the QED calculations and did not know  the  Weizs\"acker-Williams method (to the moment, mainly qualitative descriptions were spread). We started our calculations from Feynman diagrams, from the very beginning. This way allow us to skip inaccuracies widely spread in the description of similar processes even many years later.

We understand that the calculation of cross sections is more preferable than calculation of amplitudes.  Our important point was to introduce useful objects for investigation, similar to those in $ep$ DIS but more physically motivated. We find that the differential distribution  is roughly $\propto \fr{dq_1^2}{q_1^2}\fr{dq_2^2}{q_2^2}\sigma_{\ggam}^{exp}(\hat{s}, q_1^2,q_2^2)$ (more accurate form is \eqref{ggambas}) with kinematically determined lower limits\lb $q_{i, min}^2\sim m_e^2(m_e/E)^2$. In accordance with experience in the hadron physics,  we understood that  cross section $\sigma_{\ggam}^{exp}$ decreases with growth of photon virtualities like form-factor\fn{To my surprise, many physicists skip this simple fact.}. The scale of this decreasing $\Lambda$ depends on the nature of produced system.
For the most of processes of hadron production $\Lambda\sim m_\rho\sim  770$~MeV. For the production of kaons $\Lambda\sim m_\phi\sim 1$~GeV,  for the production of $\mu^+\mu^-$ pairs $\Lambda\sim m_\mu\sim 100$~MeV, for the production of discovered later charmed particles $\Lambda\sim m_\Psi\sim 3$~GeV, etc. In our estimates we  approximated $q_i^2$ dependence by step function
$\theta(\Lambda^2-q_1^2)\theta(\Lambda^2-q_2^2)$.

In the result we found that the main contribution into the cross section appears from the region of small photon virtualities $q_i^2<\Lambda^2$. In this region for the most of interesting processes the
cross section  $\sigma_{\ggam}^{exp}$ coincides with its mass shell value $\sigma_{\ggam}(\hat{s})$.
The description of differential cross sections within this region has  high accuracy  $q^2/\Lambda^2$, in the description of total cross sections it turns out to the accuracy $\sim 1/\ln(\Lambda^2/q_{min}^2)\sim 1/\ln(\Lambda^2E^2/m_e^4)\lesssim 0.03$.

We found here also estimate for  high energy total cross section\lb
\cl{$\sigma_{\gamma\gamma\to\,hadrons}\sim \sigma^2(\gamma
p)/\sigma(pp)\sim 0.3\; \mu$b,} it is in accord with
modern (2015) measurements.

Based on all these estimates, we found that the experiments at \epe colliders
{\bf open new experimental field  in the particle physics} -- the  possibility to  extract from the data an information
about  process $\pmb{\gamma^*\gamma^*\to hadrons}$,
{\bf etc},
and present necessary algorithm. We submitted our paper to the
 {\it Pis'ma ZhETF} at
May 4, 1970 and it was published there (in Russian)
 at June 5, 1970  \cite{BBG}, Fig.2, English  translation  ({\it JETP Lett})  appeared  1 or 2 months later; the abstract of this paper was published (in English) on July in the book of abstracts for XV  Rochester (in Kiev) 1970 conference (26.08-04.09) \cite{KievRoch} where paper was reported  by Budnev\fn{I cannot took part in this conference since I was in the hospital after a  car accident in the  July in Yakutia.}.
\begin{figure}[htb]
\begin{center}
\begin{fmpage}{0.73\textwidth}  \includegraphics[width=0.9\textwidth,
  height=6.5cm]{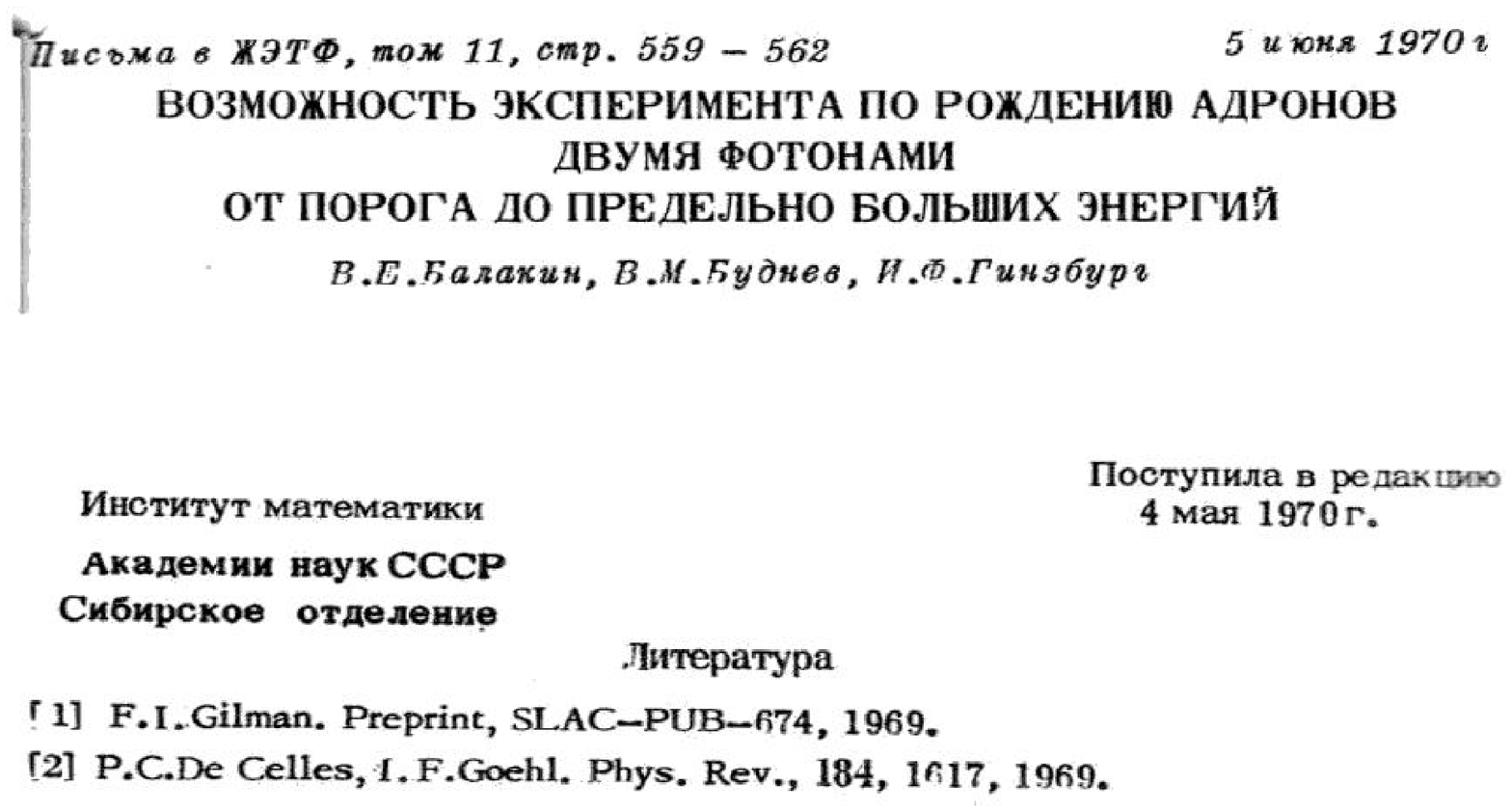}\\[-10mm]
    \caption{V.E. Balakin, V.M. Budnev, I.F. Ginzburg,
"{\it  Possible experiment of hadron production by two photons from threshold to extremely high energies
}", published June 5, 1970, submitted May 4, 1970,
Pisma Zh.Eksp.Teor.Fiz. (in Russian), transl.  in JETP Lett.}%\vspace{-5mm}
\end{fmpage}
\end{center}
\end{figure}

The paper contains also  the equations for extraction of
two-photon cross sections from the data at small electron
scattering angles in the form which is used for this aim up
to now,
 \bear{c}
\fr{d\sigma}{dE_1dE_2d\Omega_1d\Omega_2}=
\left(\fr{\alpha}{2\pi^2}\right)^2\fr{1}{q_1^2q_2^2}
\fr{E_1E_2}{E^2}\,\fr{(E^2+E_1^2)(E^2+E_2^2)}{(E-E_1)(E-E_2)}
\sigma^{\gamma\gamma}_{exp}\,,\\[4mm]
\sigma^{\gamma\gamma}_{exp}= \sigma^{\gamma\gamma}_{TT}+
\vep_1\sigma^{\gamma\gamma}_{ST}+\vep_2\sigma^{\gamma\gamma}_{TS}+
\vep_1\vep_2\left(\sigma^{\gamma\gamma}_{SS}+
\tau_{TT}\cos2\phi/2\right)
+\vep_3\tau_{TS}\,,\\[2mm]
\vep_1\!=\!\fr{2EE_1}{E^2\!+\!E_1^2},\;\;\vep_2\!=\!
\fr{2EE_2}{E^2\!+\!E_2^2},\;\; \vep_3\!=
\!\vep_1\vep_2\fr{(E\!+\!E_1)(E\!+\!E_2)}{32E\sqrt{E_1E_2}}
\cos\phi.
 \eear{ggambas}
($E$ and  $E_i$ are the energies of initial and scattered
electrons, $\phi$ is the angle between  scattering
planes of electrons, other notations was not practically changed during
45 years.) The numerical estimates of anticipated cross
sections were done and it was found that the observable cross section
grows fast with beam energy. Besides, the sketch of
experimental program was formulated. More detail calculations were published soon \cite{BBG1}.

The paper \cite{BBG} contains also the Balakin's proposal to
supplement future detectors by transverse magnetic field in the
collision region. It allows to detect the scattered forward
electrons  for the detail observation of $\epe\to \epe f$ process. This
idea was realized  on  detector MD-1 of BINP
\cite{MD_KEDR}.\\

%%%%%%%%%%%%%%%%%
\bu {\bf Brodsky, Kinoshita, Terazawa, 1970.}
%%%%%%%%%%%%
 3 month  after publication \cite{BBG}
S.~Brodsky, T.~Kinoshita \& S. Terazawa have submitted to
Physical Review
\begin{figure}[htb]
\begin{center}
\begin{fmpage}{0.85\textwidth}
  \includegraphics[width=0.95\textwidth,
  height=8.6cm]{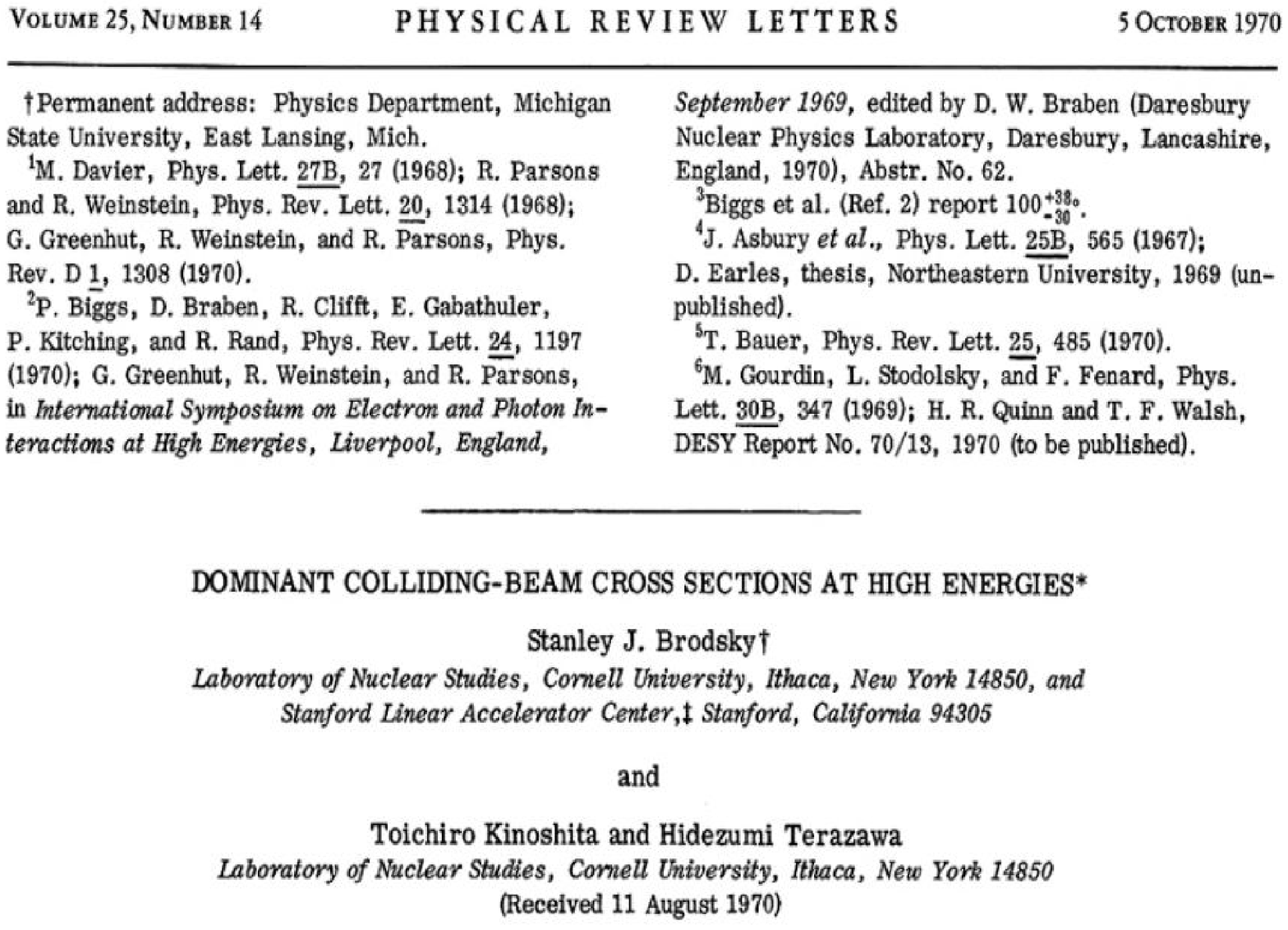}%\\[3mm]
\caption{S.J. Brodsky, T. Kinoshita \& H. Terazawa, "{\it
Dominant colliding beam cross sections at high energies}"}%\vspace{-5mm}
\end{fmpage}
\end{center}
\end{figure}
Letters their paper, published October 5, 1970 \cite{BKT} (Fig.3).
They consider two--photon production of \epe, $\mu^+\mu^-$, $\pi^0$, $\eta$ and
point-like $\pi^+\pi^-$ in \epe and $e^-e^-$ colliding
beams. They found that these cross sections grow  with
beam energy and described some features of the angular
distributions of produced pions. Analogously to \cite{BBG}, these results
had shown  that two-photon physics provides a large field
for theoretical studies and experimentation but without discussion of method of extraction information about two-photon subprocess. Unfortunately they used the Weizs\"acker--Williams method without analysis of its   applicability, with essential mistake. At the language of virtualities,
they  did not take into account the decreasing of cross sections of subprocess with virtual photons due to formfactor, and used in fact for the scale $\Lambda$, mentioned above, the kinematical limit $\Lambda\sim E$. It enhances spectra of equivalent photons by factor about 2 for each photon.   Many authors of subsequent papers reproduced this inaccuracy (I  have met  papers with such mistaken spectra even in the end of 90-th).\\

%%%%%%%%%%%%%%%%%%%%%%
\bu {\bf First experiments.}
%%%%%%%%%%%%%%%%%%%%%%%%%%%%%%%%

In 1971 VEPP-2 (BINP, Novosibirsk \cite{Bal}, \cite{KievRoch}) and in 1972
ADONE (Frascati, Italy) \cite{ADONE} reported about the
observation of $\epe\to \epe\epe$ process.\\

%%%%%%%%%%%%%%%%%%%%%%%%%%%%
\bu {\bf 1970-th.}
%%%%%%%%%%%%%%%%%%%%%

The papers \cite{BBG}, \cite{BKT}, \cite{Bal}, \cite{ADONE} open door for stream of publications devoted  to two--photon physics.

Our group continue basic analysis to understand main features of two-photon processes which are independent on the nature of produced system. In  this stage the important member of our team becomes Valery Serbo. The first results were summarized
in review \cite{BGMS} containing all necessary equations
for data preparation and set of equations useful for
different estimates. It contains also detail
description of equivalent photon (Weizs\"acker--Williams)
method, including estimate of its accuracy in different
situations. In 1974 we  did not think about possibility of longitudinal electron polarization at \epe\ storage rings and did not consider this case in basic equations. This lacuna in \cite{BGMS} was closed in \cite{GS}.

The physical problems related to the separate \ggam \ processes, details of data extraction, backgrounds  and QED processes  were discussed by many authors at that time.
Most  of papers of 70-th devoted to hadron physics in \ggam\ \  collisions were reproductions of results and ideas considered earlier for other hadronic systems.  Some of these results were reported in review \cite{BGMS}.

In 1973 series of conferences devoted  to these processes was started in Paris as the {\it International Colloquium on Photon-Photon Collisions at Electron-Positron Storage Rings}.  I cannot took part in the eight  first conferences since Soviet state  stopped my attempts despite the regular invitations. My first visit was in 1992 at  the 9-th San Diego conference from modern Russia.

The real experimental activity in this field started, in
fact, in 1979 by SLAC experiment  in which it was
demonstrated that two--photon processes can be successfully
studied at the modern detectors without recording of the
scattered electron and positrons -- via the separation of
events with the small total transverse momentum of produced
system and effective mass  $\ll  2E$ \cite{ATel79}  (this approach was initiated by V. Telnov  \cite{Teln79}).
After this work, the experimental
investigation of two--photon processes  became the essential
component of physical program at each \epe \ collider. One of the first review of
the experimental data has been summarized in the book of  Kolanoski \cite{Kolano}. Many results obtained  till now are collected  in the Particle Data Review \cite{PDG}.

At May 2 of 1980, Victor Budnev died during rafting. Since that our two-photon theoretical team from Sobolev IM-NSU consists of three  key persons -- Valery Serbo, Gleb Kotkin and me.

\section{Photon colliders }

\bu {\bf Important fact from 60-th}. In  1970 we read with great
interest about   photo-nuclear experiments in SLAC \cite{SLAC69}.
The laser photons  collided with electrons of SLAC beam producing
via backward Compton scattering  the high-energy photons. The
latter have the energy determined by production angle. Then these
{\it tagged photons} collided with  fix-target. Thus it is
appeared an opportunity to study collisions of photons having high
and precisely known energy with  proton. The typical conversion
coefficient (ratio of number of high energy photons to the number
of incident electrons) was about $10^{-7}$. The typical photon
energy was about 10\% from an initial electron energy. (After 1981,
we were informed about another similar experiments.)\\

%%%%%%%%%%%
\bu {\bf Working group at Workshop 1981. Basic idea.}
%%%%%%%%%%%%%%%%%%%%

In the winter 1980-1981, BINP was organized  the first Workshop devoted to the Linear \epe \ colliders (LC) with beam energy $E=100$~GeV, named as VLEPP.  At this Workshop I presented  the review about  two-photon physics. I concluded there that
the two-photon  option at \epe \ colliders will be  the essential part of  physical program at LC but not central point there, these studies
will give substantial supplement to the future hadron and \epe \ data with improved values of parameters but without discovery of new phenomena of the first line (except two points discussed in sect.~\ref{secphys}).

During discussions of the working group at two-photon section Valery Telnov proposed  a very new idea.
\begin{center}\begin{fmpage}{0.91\textwidth}
{\bf In the LC each electron is used only once. Therefore,  one
can try to convert
almost each electron into the high energy photon.}\end{fmpage}\end{center}
  If we can do it, one hope to obtain
\ggam and \egam collisions of real photons with luminosity about 1000 times more
than for \epe collider and with considerable  higher energies.

Unfortunately,  the  particular ideas suggested for realization of this proposal  did not look very perspective. We discussed there\ \
{\it (i)} bremsstrahlung on a solid target; {\it (ii)}
the radiation in the undulator
(wiggler); {\it (iii)}
beamstrahlung radiation in the collision with strong electromagnetic field of collided beam.

Common feature of  these proposals giving large number of produced photons was very soft energetic spectrum of these photons, large background and relatively wide angular distribution.  These  roads  had been recognized as unpromising in the discussion of the working group.

At the end of discussions  in the working group, Gleb Kotkin proposed to consider  the laser photon backscattering on the electron of LC beam in spirit of forgotten ideas of 60-th \cite{SLAC69}.  I mentioned this proposal among  others unpromising ideas in the final report of  the working group. This idea was accepted by  participants with big scepticism. They referred to the mentioned experiments (and other experiments, known to them) in which the photon energy was much lower than $E$ and the corresponding conversion coefficient was extremely small.
 Nevertheless,
Serbo and me asked Kotkin to discuss with laser experimentalists this opportunity.   In one or two days  he informed us that experts consider the necessary laser flush energy to be unacceptably high (some orders of  magnitude higher than  that of existed lasers).
{\it We were impressed that the intense high energy photon beam is only a dream.}\\

%%%%%%%%%%%%%%%%%%%%%
\bu {\bf Laser photon backscattering. First proposal.}
%%%%%%%%%%%%%%%%%%%%%%%

The idea of laser backscattering looked very attractive for us (GKS). We understood that in our case the kinematical properties of obtained photons will be better than in old works, in particular, the photons will move mainly along initial electron direction and their energy will be high enough.

Few days after Workshop during our walking with Serbo I suggested: "Let us check statements of Gleb" (we know that he  may be impressed by the opinion of a good person  and give up after the first objection).   During walk we  estimated necessary laser flash energy. Our estimate was very simple. We were known the size of electron beam of VLEPP near the collision point $S$. For complete conversion of electrons to photons the laser target should be opaque  for electrons. Therefore, the necessary number $N$ of laser photons in flash is $S/\sigma_C$,  where $\sigma_c$ is Compton cross section. For the first oral estimate we  took  for $\sigma_C$  the Tomson limit value. For the laser photon energy $\omega_o \sim$ few eV (visible light) we estimated  the necessary laser flash energy $\omega_o N\sim  10$~J. This value seemed realistic for us. One half hour later Serbo at home reproduced this estimate with pen.

After that we three  were connected with laser specialist Folin about possible type of laser suitable for our problem. He showed us laser from neodimium glass or garnet with laser photon energy $\omega_o=1.17$~eV ({\it this choice turned out to be the best up to now}). He informed us about existence such lasers with  the necessary flash energy and   repetition rate (about 100~Hz) -- separately.  He told us that with suitable budget even middle laser group can construct laser with necessary flash energy and repetition rate for  about 3 year. We understood that the desirable conversion can be possible.

\begin{figure}[htb]
\begin{fmpage}{0.45\textwidth}
\includegraphics[width=0.98\textwidth,height=4.5cm]{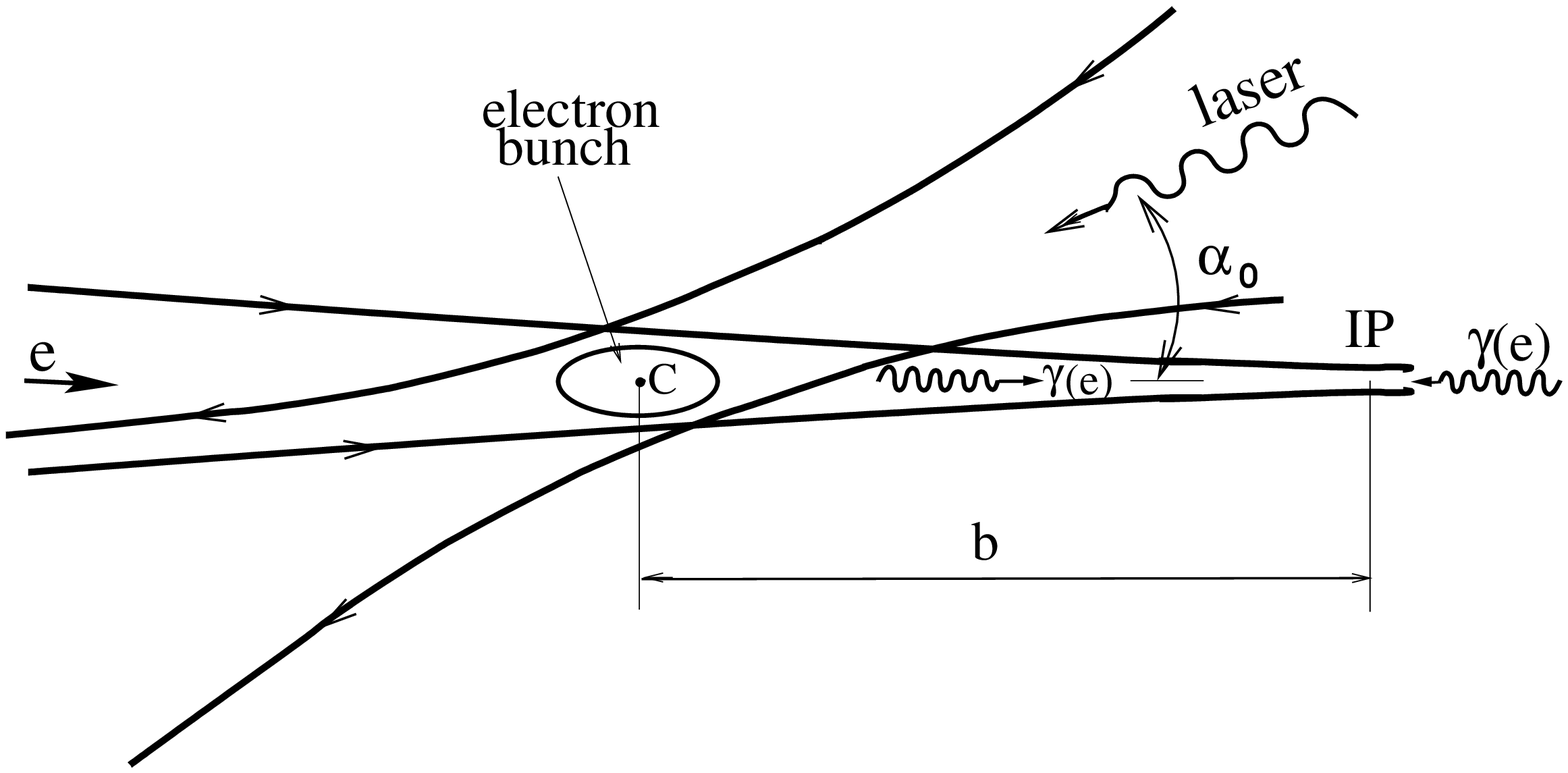}\vspace{-4mm}
\caption{\it  Scheme of   conversion (figure from \cite{Telnov})}
\end{fmpage}  \label{fig:basschem}\hspace{5mm}
\begin{fmpage}{0.4\textwidth}
\includegraphics[width=0.96\textwidth,height=4.5cm]{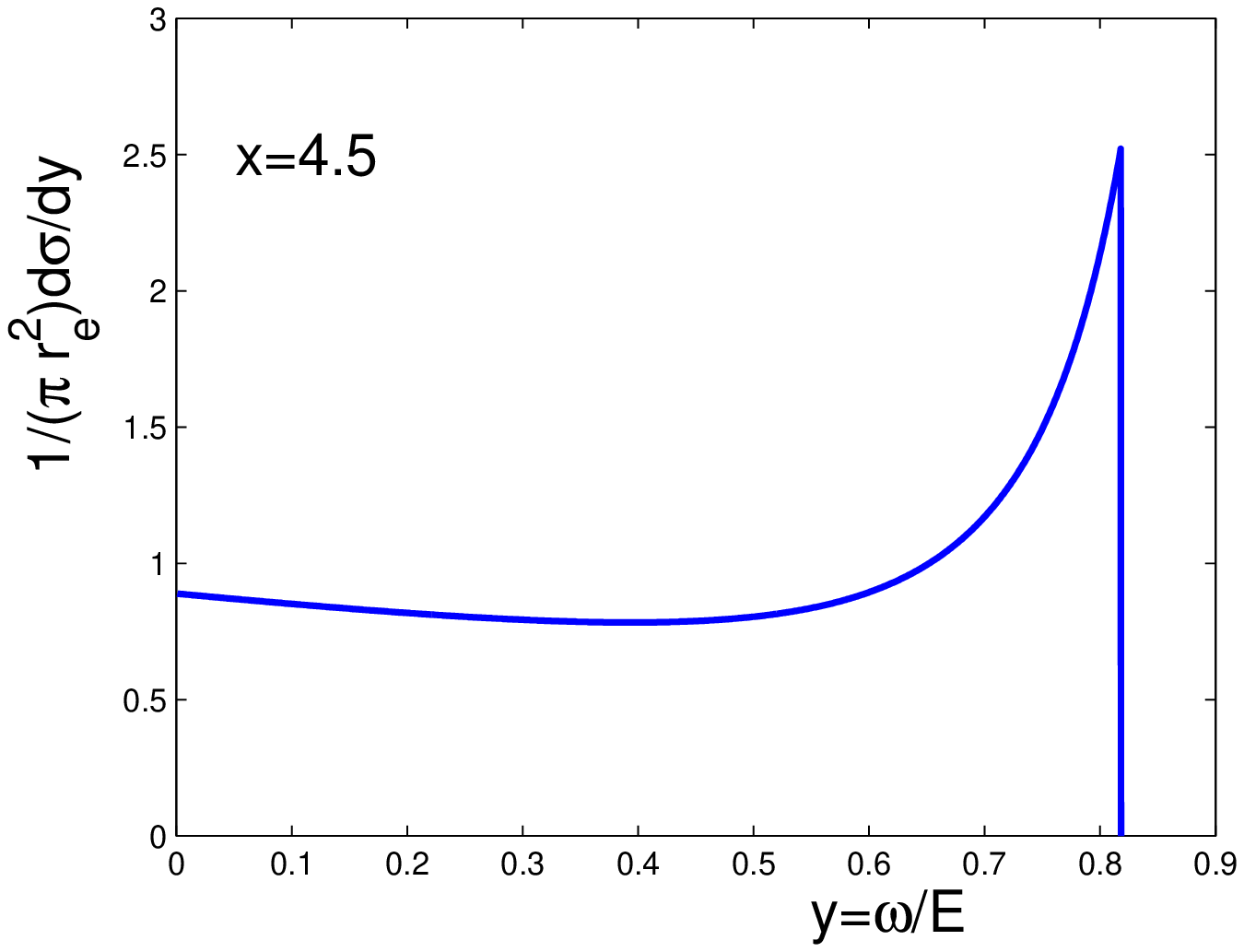}\vspace{-4mm}
\caption{\it Energy spectrum for unpolarized photons, $x=4.5$} \end{fmpage}\label{fig:basspectr}
\end{figure}

The scheme of $e\to\gamma$ conversion was evident for us  from the very beginning (Fig.~4). At~the conversion
point $C$, preceding  the interaction point $IP$,  the
 electron ($e^-$ or $e^+$) beam of basic Linear Collider (LC)
meets  the photon flash from powerful laser containing photons $\gamma_o$, having energies $\omega_o$ (We neglect below difference of the angle $\alpha_0$ in
Fig.~4 from 0.). The Compton
backscattering of laser photons on electrons from LC $\gamma_oe\to\gamma e$ produces high
energy photons $\gamma$, having energies $\omega$, with energy spectrum limited by the kinematically determined upper limit $\sim E$. With   the suitable choice of laser one can
obtain the photon beam with the photon energy
close to that of the basic electron.
 The ratio of number of high energy photons to that of electrons --
the conversion coefficient $k\sim 1$.

To describe phenomenon we introduce variables $x=4E\omega_o/m_e^2$ and\lb $y=\omega/E$, so the squared Compton cms energy $\hat{s}_C=(x+1)m_e^2$. Simple kinematic calculation showed that $y$ is limited from above by quantity $y_m=x/(x+1)$. Using the well known QED results for the considered case,  we found that the energy spectrum of photons is concentrated near upper  limit  $y_m$, Fig.~5.

We were lucky with the numbers. Indeed, at the considered electron energy $E=100$~GeV the parameter $x\approx 1.8$.
Therefore, the maximal photon energy  is equal to $\omega_m=0.64E$ (while at earlier experiments, for example,  at $E=10$~GeV we had
$x=0.18$ and $\omega_m=0.15E$, as what mentioned by participants of Workshop. The  choice of lasers with reasonable high power flash allows in principle to reach conversion coefficient $\sim 1$, in contrast with $10^{-7}$ in experiments \cite{SLAC69}.

Since cm energy of $\gamma_oe$ system is $m_e\sqrt{x+1}$, the
transverse momentum of produced photon is $\lesssim
m_e\sqrt{x}\lesssim 1$~MeV, and  the photon escape angle $\theta $
is typically very small ($\sim m_e/E$)  and depends on the escape
angle as $y=y_m/(1+(\theta/\theta_0)^2)$, where
$\theta_0=m_e\sqrt{x+1}/E\lesssim 10^{-5}$. Therefore,  the
produced photons move almost along momenta of incident electrons
and focus  approximately in the same   spot, as it is expected for
electrons without laser conversion. Hence, the total luminosity
provided by these photons will be close to that expected for
initial \epe \ or $e^-e^-$ collision. And  the repetition
rate for VLEPP project (100 Hz) seemed realistic for laser
community.

In the IP the obtained photon beam collides with either opposite non-converted electron beam (\egam \ collisions) or with photon beam (\ggam \ collisions). Later on this scheme was called {\bf Photon Linear Collider} --- PLC.

In few days we reach  complete understanding. At this stage  Valery Telnov    joined us. It  became  clear that the opportunity is realizable and the paper with  the
corresponding proposal should be written as soon as possible.
Our joint studies  were published in Refs.~\cite{GKST81a,GKST81b,GKST83}.

We develop together the concept of the differential luminosity spectrum. It is given by convolution of individual photon spectra with geometric factor determined by the  energy dependent angular spread. At the
shift of conversion point C from the interaction point IP  on the distance $b$ (Fig.~4), the photons of smaller energies spread for more wide region, and their contribution
into luminosity become relatively lower, i.e. total luminosity decreases, but lumi-
\begin{wrapfigure}[14]{l}[0pt]{0.37\textwidth}%\vspace{-.5cm}
\begin{fmpage}{0.37\textwidth}
\includegraphics[width=0.98\textwidth]{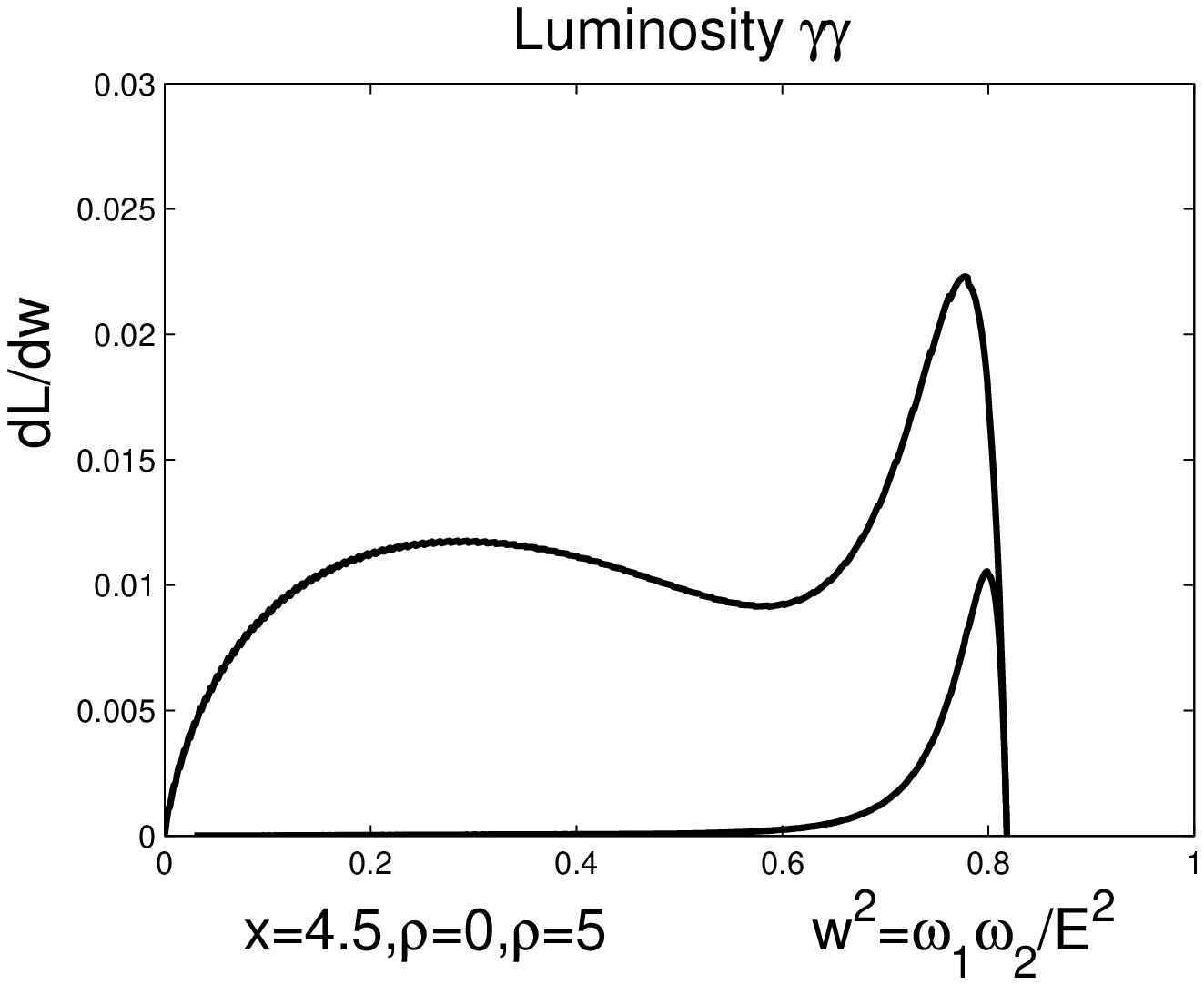}\vspace{-3.5mm}
\caption{\it Luminosity spectra at  $\rho=0$ and 5 for unpolarized photons, $x=4.5$} \label{fig:baslum45}
\end{fmpage}
\end{wrapfigure}
nosity spectra become more monochromatic -- quality of \ggam \ collisions  improves,
Fig.~\ref{fig:baslum45} ({\it  L.~Barkov suggested us to underline this fact.}).

The  dependence of luminosity spectrum  on the distance $b$ for the case of the round electron beam  is determined by  parameter, expressed via the  radius of this beam $\sigma$ in the IP
for the case without conversion:
\be
\rho=bm_e/(\sigma E)\,. \label{rhoround}
 \ee

\bu The important  remaining problem  is the following: The length
of electron bunch is finite. Within this length it is necessary to
provide the density of laser photons sufficient for conversion. We
understood that the density of laser flash  decreases with the
growth of distance from the focal plane. Kotkin suggested to use
the Gaussian laser beams providing the maximal length of region of
highest density of photons. Simple estimate gave  the optimistic
result. Serbo confirmed it by direct calculation with these beams.
It transformed  our preliminary estimates into reliable
calculation.

In these papers we suggested to remove residual electrons from IP by magnetic field $\sim 1$~Tl, acting between conversion point and IP. 10 years later    Telnov  \cite{Telnov}, \cite{Teln2015}  offered to  abandon the use of the magnetic field.

\bu \ We note in this basic paper that the quality of \ggam \ collisions (degree of monochromaticity) improves with increase of $x$. However, with the growth of $x$ the new phenomenon stops improvements. At large enough $x$ the number of output high energy photons is diminishes due to their death  in the collisions with laser  photons from the tail of laser flash, produced  \epe \ pairs. Therefore, the "optimal" laser photon energy  is limited from above by the threshold of $\gamma_0\gamma\to \epe$ process $x>2(1+\sqrt{2})\approx4.8$ (for the considered laser at $E=250$~GeV we have almost optimal $x=4.5$).

In two or three months after sending of preprint \cite{GKST81a} in different centers including SLAC, we receive preprint \cite{Akerlof} with similar in form proposal for \egam \ collisions but with incorrect estimate of necessary laser flash energy. After our message pointing this inaccuracy  author stops his activity.

The first journal publication \cite{GKST81b} meet unexpected objection of deputy editor of JETP Letters -- "publication is unsuitable since necessary lasers don't exist now". The overcome of this  unfair objection delays publication for 4 or 5 months.\\

%%%%%%%%%%%%%%%%%%%%%%%%%%%%%
\bu {\bf Polarization.}
%%%%%%%%%%%%%%%%%%%%%%%%%%%%%%%

During Workshop we have learned about the  possibility to obtain longitudinally polarized electrons in the project of LC. Laser light is easily polarized.

We known that the polarization effects can give only little changes for high energy hadron and \epe \ collisions. We believed that the same is also true for the $\gamma \gamma$ collisions. As a result, the study of such effects at the $\gamma \gamma$ collisions looks for us as useful but not important problem.

Nevertheless, our theory group (GKS) started  to calculate the polarization effects. The first analysis  gave us the surprising result. {\it The energy spectrum of photons changes strongly for longitudinally polarized collided particles}. This spectrum depends on  the
product of longitudinal polarization of electron (helicity) $\lambda_e$ and degree of laser circular polarization $\lambda_L$.    At $2\lambda_e\lambda_L=-1$ the number of photons with  the maximal energy is almost doubled as compare with the case of nonpolarized photons. On the other hand,  at $2\lambda_e\lambda_L=1$ photons with the maximal energy almost disappear  (see Fig.~7).
\begin{figure}[htb]
\begin{fmpage}{0.45\textwidth}
\includegraphics[width=0.98\textwidth]{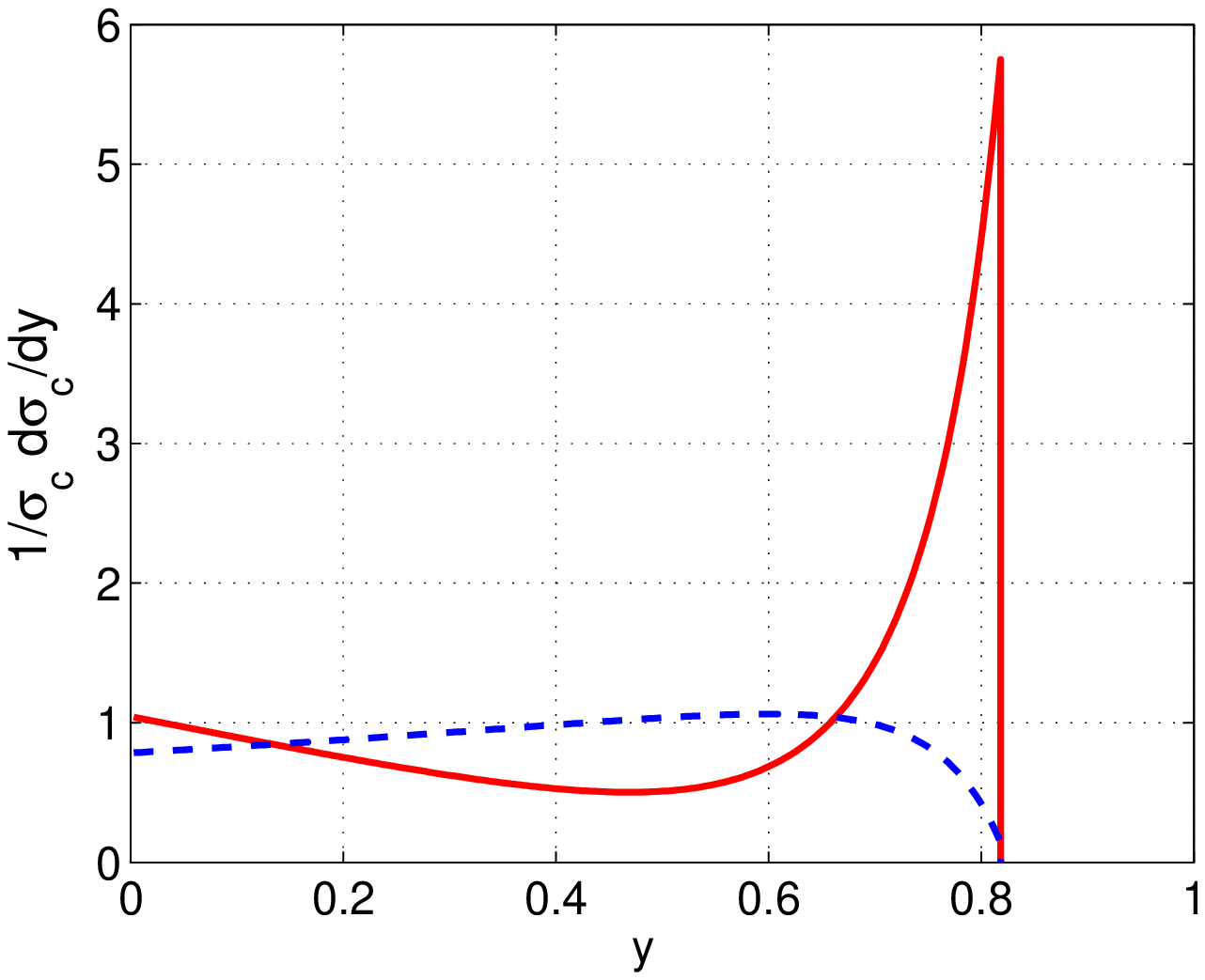}\vspace{-4mm}
\caption{\it Photon energy spectrum  at $x=4.5$, $2\lambda_e\lambda_L=-1$ (full)
and $2\lambda_e\lambda_L=1$ (dotted)}
\end{fmpage} \label{fig:Compsp48}\hspace{5mm}
\begin{fmpage}{0.45\textwidth}
\includegraphics[width=0.97\textwidth]{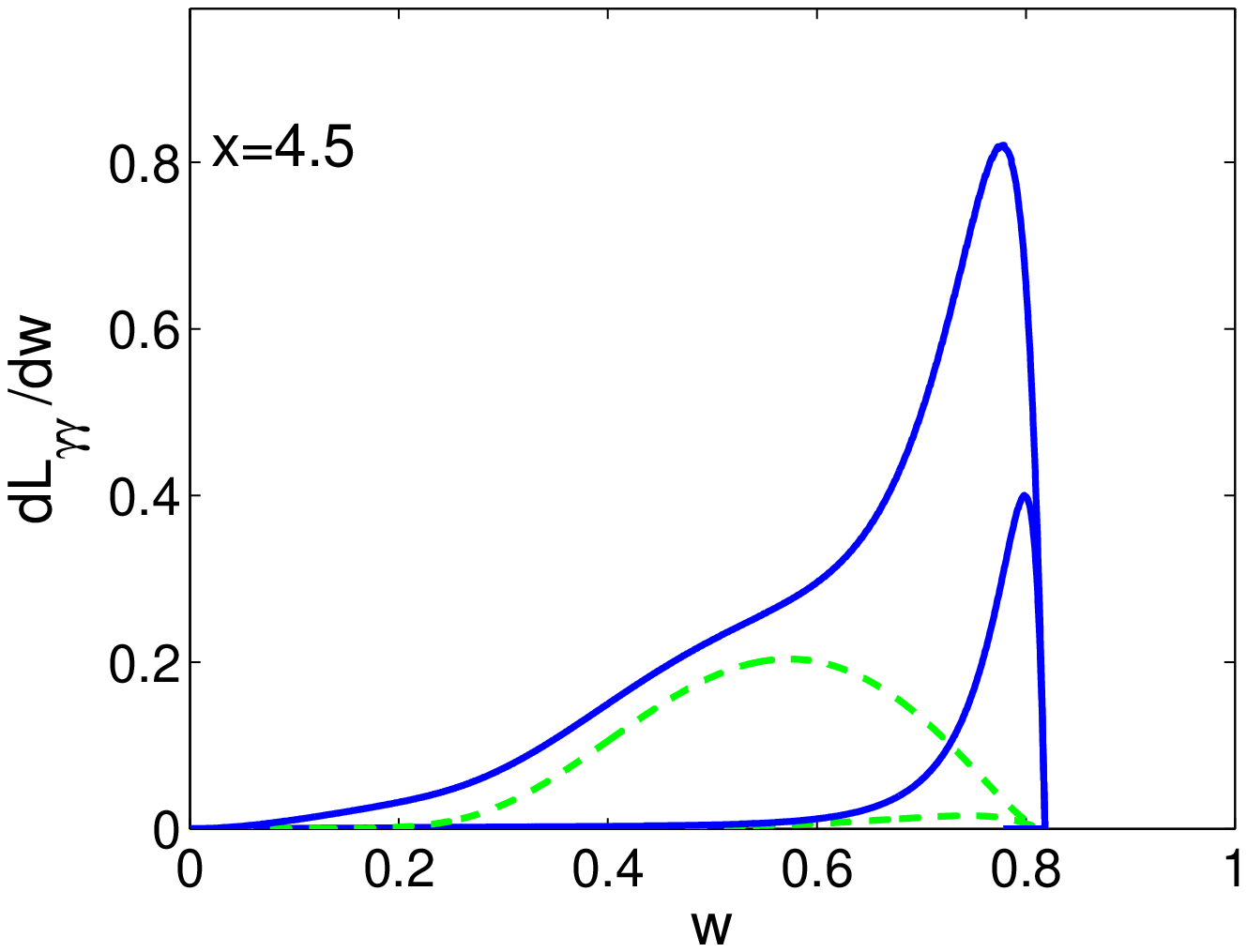}\vspace{-4mm}
\caption{\it Luminosity spectra  at $x=4.5$, $2\lambda_e\lambda_L=-1$ -- $L_o$ (full) and $L=2$ (dotted)
for $\rho=1$ (upper) and $\rho=5$ (lower)}
\end{fmpage} \label{fig:Lggampol48}
\end{figure}
(The total cross section depends on polarization weakly
at $x\lesssim 8$).

This observation was the reason for more detail study of problem.
First, we observed that the circular polarization of laser photons
is transferred to that of high energy photons $\lambda$. At
$y=y_m$ we have $\lambda=-\lambda_L$  due to angular momentum
conservation. At smaller $y$  value of $\lambda$ decreases,
depending on both $y$ and the parameter $2\lambda_e\lambda_L$ for
the incident beam. Therefore, it is useful to consider two
different  $\gamma\gamma$ luminosities, dependent on the initial
laser photon polarization -- the luminosity $L_0$ for  high-energy
photons having identical helicity (total helicity of final state
 $\lambda_1-\lambda_2=0$) and $L_2$ for photons having opposite
helicity (total helicity of final state
$|\lambda_1-\lambda_2|= 2$). At the suitable choice of initial
helicities $L_0>L_2$  (see Fig.~8).

The natural next problem was to study transverse polarization of photons. To the moment we did not know detail equation for Compton effect with all polarization. Besides, at the subsequent stage we meet important technical difficulty. The scattering planes, which are useful for description of transverse polarizations in the individual Compton process, are different for each  Compton scattering. In the description of beam polarization  the suitable averaging becomes necessary. Some delicate effects appear at this averaging.   My recent PhD student Shimon Panfil  took part in the  corresponding calculations.  The results were published in \cite{GKPS83gg}. These  complete description of polarization phenomena  was presented in the paper prepared together with V.~Telnov \cite{GKPST84}.

Let us summarize  some results of this study for collision of longitudinally polarized electron beam and polarized laser beam. The photons with  the highest energy $y\sim y_m$ can be made circularly polarized with high degree of polarization using the circularly polarized laser light. Degree of this polarization increases with the growth of $\rho$.  The sizable transverse polarization of high energy photons can be obtained using the transversally polarized laser light, but only at moderate $x<2$ and $y\lesssim y_m/2$.
Degree of this polarization  is not high  and it decreases with the growth of $\rho$.

The papers \cite{GKST83} and  \cite{GKPS83gg}, \cite{GKPST84} gave the complete description of basics of PLC.
Beginning from 90-th many  physicists considered different problems related  to construction of PLC, but in fact  these investigations added some details which changed  basic results only weakly.

After these basic researches, our team (GKS)  studied  the
physical processes at PLC with some works devoted to PLC itself,
while Telnov concentrated efforts on the technical problems
\cite{Teln2015}.  His activity in the various audiences ensured the
inclusion of PLC mode in all projects of LC's. The challenges for
the PLC project  were given by strong increasing of repetition
rate as compare with  the original VLEPP project, strong
decreasing of beam size of LC and corresponding electromagnetic
field of laser bunch, choice of geometry of collision, etc. Telnov
answered to the most of these challenges with the goal to obtain
the highest \ggam \ luminosity of the best quality.

 Beginning from 90-th Photon Collider become substantial part of Linear Collider projects \cite{TESLA}, \cite{ILC}.\\

%%%%%%%%%%%%%%%%%%%%%%%%%%%
\bu {\bf The attempt to use  infrared lasers. Nonlinear QED effects.}
%%%%%%%%%%%%%%%%%%%%%%%%%%
The next problem interested for us after first studies was
the following.

At $E=100$~GeV we have $x\approx 1.8$ which is far from optimal
value $x\approx 4.8$. {\it How to obtain photons with higher
energy?}

We did not know about perspectives to construct powerful laser
with photon energy about $2.5\div 3$~eV which give necessary $x$.
The using of  a free electron laser with
the regulated frequency demands one more complex
equipment. However, another idea looks attractive for the first
glance.

We  knew about  the  very powerful gaseous laser with
$\omega_o\approx 0.2$~eV  (e.g. on $CO_2$) and   good repetition rate. Therefore, it seemed attractive
to use such laser with the very high flash energy.  In this case
the laser photon density will be so high that the typical process
will be not the ordinary Compton scattering, but the collision of an
electron with a few laser photons simultaneously. It will be the
non-linear QED (NQED) process like $e+5\gamma_o\to \gamma +e$ (see
\cite{Ritus} for  the basic description). Our first naive
expectation  was that in this way one can reach necessary high
energy of  the final photons. Kotkin,  my  PhD student Polityko
and me consider this problem\fn{Simultaneously we consider
production of \epe \ pairs in the NQED processes like
$e+5\gamma_o\to \epe +e$. This process can be used as the signal of observation of NQED
processes \cite{GKPel}. } \cite{GKPphot}. We found that the
desirable parameters  can be obtained on the existent lasers.
Nevertheless,  the real situation appeared far from our
expectations due to new effects  in the strong electromagnetic
field. These effects are determined by the parameter
 \be
\xi^2=\fr{e^2F^2}{(mc\omega_o)^2}\equiv
\fr{4\pi\alpha\hbar}{m^2c\omega_o}n_L\,.
\label{nelparam}
 \ee
Here $F$ is electric field strength in  the laser wave and
$n_L$ is the laser photon density in the conversion
region. At low and moderate $\xi$ the probability of process with
simultaneous collision of $k$ laser photons $e+k\gamma_o\to \gamma
e$ is proportional to $(\xi^2)^k$ (with small numerical factor).
However,  the transverse motion of electron enhances its effective mass as $m^2\to (m^*)^2=m^2(1+\xi^2)$. The  maximal photon
energy decreases as $y_m=kx/(1+kx+\xi^2)$. At $\xi\gg 1$  the new parameter $\chi=x\xi$
becomes important. At $\chi\gg 1$ the energy
distribution of produced photons is roughly
similar to that for virtual photons \cite{GKPphot}.

As a result, at any $\xi$ the  fraction of photons with really
high energy will be very low, therefore, this
approach is unsuitable for construction of PLC.

20 years  later D. Ivanov, G. Kotkin and V. Serbo presented the
complete description all polarization effects in the non-linear
Compton scattering which is used now for simulation of high-energy
photon production in the conversion region \cite{IKP}.\\

\bu {\bf Measuring and simulation of luminosity.}
%%%%%%%%%%%%%%%%%%%%%%%%%%%%%%%%%%%%%%%%%%%%%%%%%%%

In the  papers \cite{GKST81a}-\cite{GKST83}, \cite{GKPST84} we noted that the future luminosity distributions will differ from those, calculated in our papers, and experimental calibration is necessary. We had in mind corrections obliged by non-round geometrical form of electron bunches, rescatterings and other  processes in the conversion  and collision regions. We also think about inaccuracy in the aiming of beams. Telnov present simulation of all effects {\it except inaccuracy in the aiming} and to publish "realistic spectra of luminosity" \cite{Telnovspectr}. The method for measuring this distribution via processes $\ggam\to \epe, \mu^+\mu^-, \epe\gamma, \mu^+\mu^-\gamma, \epe\mu^+\mu^-$ was presented by Serbo, Telnov et al. in  \cite{ggamlum}.\\

%%%%%%%%%%%%%%%%%%%%%
\bu {\bf Flat electron beams.}
%%%%%%%%%%%%%%%%%%%%%%%%%%

In modern projects the  electron beam for LC will be very flat (in the {\it would be}
interaction point this beam presents ellipse with half-axes
$\sigma_x$ and $\sigma_y$, where $\sigma_x/\sigma_y\lesssim 100$).
 It leads to   changes in the
luminosity spectra for \ggam collisions, especially in low energy
part. In the paper \cite{GKrho} Kotkin and me consider the high
energy part of these spectra  (which in addition  weakly changes by
rescatterings). We found that  this high energy part with good accuracy is described by the same
equation as for the round beam with the natural replacement of
\eqref{rhoround} to \be
\rho^2=\left(b/(E/m_e)\sigma_x\right)^2+\left(b/(E/m_e)\sigma_y\right)^2\,.
\label{rhodef}
 \ee

%%%%%%%%%%%%%%%%%%
\bu {\bf The case $x>4.8$.}

The most suitable modern lasers with neodymium glass or
garnet allow to realize  the basic scheme in its pure
form only for the electron  energy $E\le 250$~GeV (the first stage of ILC). At $x>2(1+\sqrt{2})\approx 4.8$ (at
$E>270$~GeV with the same laser), some of produced high energy
photons are died out, producing \epe pairs in the collisions with
laser photons from the tail of laser bunch. This fact was treated
as limiting one for realization of PLC based on LC with
higher electron energy \cite{GKST81a}-\cite{GKST83}).

The opportunity to use standard scheme at $x>4.8$ with lower conversion coefficient
 was mention in
\cite{Tel2001,GKPho9} but without detailed description. In the
ref.~\cite{GinKot15} Kotkin and me  found that  using  the same laser system with almost the same laser
flash energy  as was prepared for
$E=250$~GeV allows to obtain PLC with  $E\le 1$~TeV and
luminosity concentrated in the high energy part only, with $\Delta
\hat{s}/\la\hat{s}\ra\sim 3\div 5\%$. The total luminosity at
suitable $\rho$ is $0.25\div 0.2$ from that for high energy part
of luminosity at $x=4.5$ ($E=250$~GeV).\\

%%%%%%%%%%%%%%%
\bu {\bf Change of photon polarization. Vacuum birefringence.}
%%%%%%%%%%%%%%%%%%

The scattering of high energy photon on laser photon from the tail
of laser bunch after conversion can result in  an
interesting effect, considered by Kotkin and Serbo
\cite{Kotserpolar}. At $x>4.8$ this collision results in
production of \epe \ pairs, discussed above.

At $x<4.8$ the main interaction is the elastic \ggam \ scattering with
negligibly small cross section ($<\alpha^4/m_e^2$). Therefore, the
laser bunch is practically transparent for such $\gamma$-quanta.
On the other hand, the variation in polarization for the
$\gamma$-quantum traversing the bunch is determined by the
interference of the incoming wave and the wave scattered at zero
angle. In other words, for such a variation it is responsible not
the cross section (which is proportional to square of the
light-light scattering amplitude of the order of $\alpha^4$), but
the scattering amplitude itself $\sim \alpha^2$. As a result, in
this case the essential variation in the $\gamma$-quantum
polarization can occur practically without loss in intensity of
$\gamma$-quanta --  {\it vacuum birefringence}. Fortunately, this effect
weakly influences for the photons of highest energies for the case
with longitudinally polarized incident beams. However, it should
be taken into account at intermediate $y$ and for transverse
polarization.

%%%%%%%%%%%%%%%%%%%%%%%%%%%%%%%%%%%%%%%%%%%%
\section{Notes on physical program}\label{secphys}
%%%%%%%%%%%%%%%%%%%%%%%%%%%%%%%%%%%%%%%%%%%%%%

The physical program for photon collisions has huge literature.

\bu \ {\bf $\pmb {e^\pm e^-}$ colliders}. Energies and
luminosities of \ggam \ subprocesses are  much lower  than
those of parent colliders. Therefore, the two-photon studies at
these colliders provide  substantial supplement to the future
hadron and \epe \ data with improved values of parameters but
without discovery of new phenomena of the first line.

\bu {\bf The photon colliders} will be built only in far future. No
doubts, measurements at these colliders will  improve accuracy of
results obtained at hadron and \epe \ colliders. I skip
these problems here.

 Below I discuss only several processes in which two-photon mechanism can provide
information unavailable in other collisions or machines. For \epe
\ collisions that are points (A) and (B) below, for PLC -- points
(B)-(G).

(A) At relatively low \ggam \ energy the interference between
two-photon and bremsstrahlung mechanism of production of simple
systems like $\pi^+\pi^-$ allows  \underline{to measure}
relative phases of s- and p-waves (d- and p-waves) of $\pi\pi$
scattering, not available in other approaches \cite{SerCher}.

(B) The most important for photon collisions  at \epe \ colliders
and very important for PLC is the study of the structure function
of the photon. Witten found that it is an unique quantity in
particle physics which can be determined from QCD at large enough
$Q^2$ and $s$ completely without phenomenological parameters, it
is determined by point-like component of photon \cite{Witten}. The
 test of this result  at future
experiments is necessary to verify that QCD is indeed a theory of
strong interactions.  At modern parameters of \epe \ machines the hadron-like component of
photon dominates.

\begin{center}
\begin{figure}[htb]\centering
\begin{fmpage}{0.7\textwidth}
\includegraphics[width=0.98\textwidth,height=0.4\textheight]{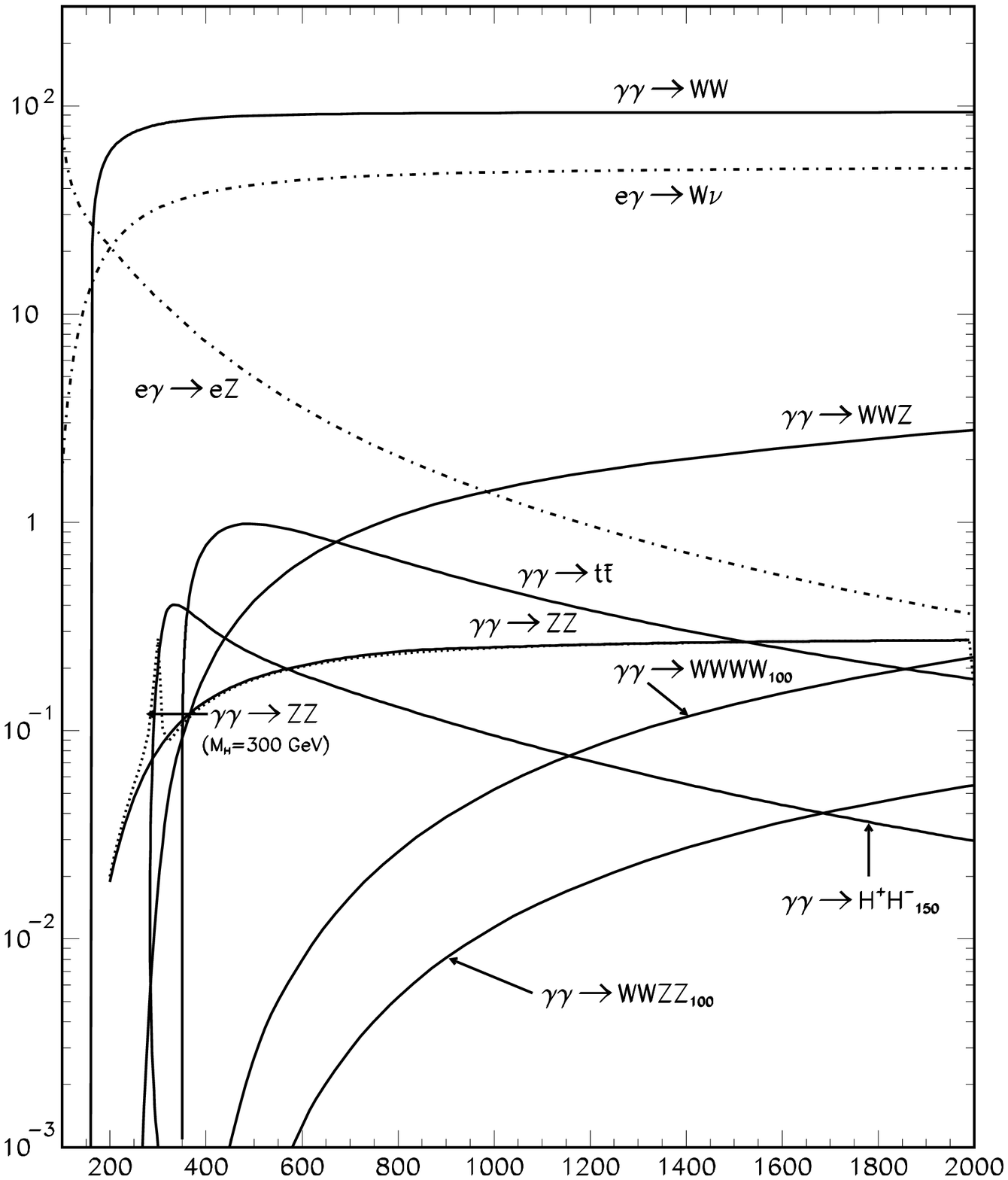}\vspace{-4mm}
\caption{\it Cross sections of some processes at PLC in {\it pb}. Unpolarized photons. Subscript $_{100}$ means that it is calculated for $m_h=100$~GeV}
\end{fmpage} \label{fig:processes}
\end{figure}%\end{center}
\end{center}

(C) The study of semi-hard processes like $\ggam\to
\rho\rho$ provides an opportunity to measure  the
Pomeron of QCD with high accuracy. We presented  one of the
first calculations in this field \cite{GPS87}. Now
this  topic has huge literature.

(D) The next group form processes with production of gauge $W$
and $Z$ bosons.  In 1983 we consider the basic processes of gauge
boson production in $\gamma \gamma$ and $\gamma e$ collisions
\cite{WGKPS83}. The important observation was that the cross
sections of these processes don't decrease with the growth of
collision energy, reaching large enough value $\sigma_W\sim 80\div
90$~pb. (That is due to dominance of vector  -- $W/Z$ -- exchange
in $t$-channel). It allows to perform the high energy measurements
with high enough accuracy. It opens  the door for the test of high
order electroweak radiative corrections, which cannot be studied
at another machines with reasonable accuracy. The relevant field
provides an opportunity to observe  the multiple production of
gauge bosons in processes like $\ggam \to WWZ$, $\egam\to eWW$,
$\egam \to WZ\nu$ with cross sections $\sim \alpha\sigma_W$ and
$\ggam\to WWWW$, $\ggam\to WWZZ$, $\egam\to eWWZ$, $\egam\to\nu
WWW$ with cross sections $\sim \alpha^2\sigma_W$ \cite{GinIl93} (see Fig.~\ref{fig:processes}).

(E) The  study of
resonances with spin 0 or 2 in the processes $\ggam\to WW$,
$\ggam\to ZZ$ which can appear due to possible strong interaction
in Higgs sector allowed by modern data in the multi-Higgs models.

(F) The study of Higgs productions at PLC has huge literature.
The probing of possible violation of CP in the extended models of Higgs sector is very difficult task for LHC. It can be solved in the study of process $\ggam\to h$ with polarized photons \cite{ZerGin}.

Unfortunately, we can hope now that these measurements will give only refinements of data obtainable at LHC and future
LC. I don't expect that these refinements would be crucial in the
understanding of general picture. Moreover, observations of
production of possible heavier neutral Higgs bosons at PLC look
difficult task \cite{Gin2015b}.

(G) Who knows?

\section*{ Acknowledgments}

 I am thankful  to my friends Valery Serbo and Gleb Kotkin,
my co-authors in the most of discussed papers, for many useful
comments and support of idea to present this report. I am thankful
Valery Telnov for long time fruitful collaboration in Photon
Collider project and discussions.
Special thanks to D.V. Shirkov
 after conversation with whom I decided
 to publish this report.

This work was supported in part
by grants RFBR 15-02-05868, NSh-3003.2014.2 and  NCN OPUS 2012/05/B/ST2/03306 (2012-2016).


\section*{References}



\begin{thebibliography}{99}

\bibitem{LL} L.D. Landau, E.M. Lifshitz, {\it Sow. Phys.}
{\bf 6} (1934) 244

\bibitem{BH} H.A. Bethe, W. Heitler, {\it Proc. Roy. Soc.}
{\bf A 146} (1934) 85

\bibitem{Racah} G. Racah, {\it Nuovo Cim.} {\bf 14} (1937)
93.

\bibitem{LRum} H. Baba, W.A. Heitler, {\it Proc. Roy. Soc.}
{\bf A 159} (1937) 432; J. Carlson, J.R. Oppenheimer, {\it Phys.
Rev.} {\bf 51} (1937) 220; L.D. Landau, G.B. Rumer, {\it
Nature} {\bf 140} (1937) 682; {\it Proc. Roy. Soc.}{\bf A
196} (1938) 213.

\bibitem{enloss} S.R. Kelner, {\it Sow. Yad. Fiz.} {\bf 5}
(1967) 1092; S.R. Kelner, Yu.D. Kotov, {\it Sow. Yad. Fiz.}
{\bf 7} (1968) 360; G. Khristiansen et al., {\it Sow. Yad.
Fiz.} {\bf 15} (1972) 966.

\bibitem{Prim} H. Primakoff, {\it Phys. Rev.} {\bf 81}
(1951) 889; A. Halperin et al. {\it Phys. Rev.} {\bf 152}
(1966) 1295.

\bibitem{Low} F.E. Low, {\it Phys. Rev.} {\bf 120} (1960)
582.

\bibitem{CalZem} F. Calogero, C. Zemach, {\it Phys. Rev.}
{\bf 120} (1960) 1860.

\bibitem{deCel} P.C. De Celles, J.F. Goehl, {\it Phys. Rev.} {\bf
184} (1969) 1617.

\bibitem{Kess} N. Arteaga-Romero, A. Jaccarini, P. Kessler,
{\it Compt. Rend.} {\bf 269B} (1969) 153, 1129;
A.~Jaccarini et al., {\it Lett, Nuovo Cim.} {\bf 4} (1971)
933; N. Arteaga-Romero et al. {\it Phys. Rev.} {\bf D3}
(1971) 1569, 1927.

\bibitem{Bai} V.N. Baier, V.S. Fadin, {\it Lett. Nuovo Cim.} {\bf 1}
(1971) 481

\bibitem{Bal} V.E. Balakin et al. {\it Phys. Lett.} {\bf 34
B} (1971) 320; {\it Yad. Fiz.} {\bf 16} (1972) 729; E. Pakhtusova, report on this Conference.



\bibitem{KievRoch} XV Int.Conf. on High Energy Phys., Rochester-Kiev. Book of abstracts, Kiev (1970)

\bibitem{BBG} V.E. Balakin, V.M. Budnev, I.F. Ginzburg, {\it
Pis'ma ZhetF} {\bf 11} (1970) 559 ({\it ZhETF Lett.} {\bf
11} (1970) 388).


\bibitem{BBG1}
V.M.~Budnev, I.F.~Ginzburg, {\it Phys. Lett.} 37B (1971) 320; {\it Sov. Yad. Fiz.} 13 (1971) 353.


\bibitem{MD_KEDR} S.E.Baru et al., {\it Phys. Rept.} {\bf 267} (1996) 71.


\bibitem{BKT} S. Brodsky, T. Kinoshita, H. Terazawa, {\it
Phys. Rev. Lett.} {\bf 25} (1970) 972.


\bibitem{ADONE} C. Bacci et al. {\it Lett. Nuovo Cim.}, {\bf
3} (1972) 709; G. Barbellini et al., {\it Phys. Rev.
Lett.}, {\bf 32} (1974) 385.



\bibitem{BGMS}
V.M.~Budnev, I.F.~Ginzburg, G.V.~Meledin and  V.G.~Serbo,
{\it Phys. Rep.} {\bf 15C} (1975) 181.
\bibitem{GS}  I.F. Ginzburg, V.G. Serbo. {\it Phys.\ Lett.}
{\bf B 96} (1980) 68.






\bibitem{ATel79} G.~Abrams et al., {\it Phys. Rev. Lett.} {\bf 43} (1979) 477
\bibitem{Teln79} V.I. Telnov, Report on this Conference

\bibitem{Kolano} H. Kolanoski, {\it Two photon physics at \epe \ storage rings}.
    Springer Tracts in Mod. Phys. {\bf 105} (1984) 1-187
    BONN-HE-84-06



\bibitem{PDG} Particle Data Group. {\it Chinese Phys.} {\bf C
38} (2014) 1-1661


\bibitem{SLAC69}
J.~Ballam  et  al., {\it Phys. Rev. Lett.} {\bf 23} (1969) 498.




\bibitem{GKST81a}
I.F.~Ginzburg, G.L.~Kotkin, V.G.~Serbo, V.I.~Telnov,
{\it Preprint 81-50 BINP}.  (May 1981);
{\it Pis'ma ZhETF} {\bf 34} (1981) 514.

\bibitem{GKST81b}
I.F.~Ginzburg, G.L.~Kotkin, V.G.~Serbo, V.I.~Telnov,
{\it preprint 81-102 BINP}  (1981).

\bibitem{GKST83}
I.F.~Ginzburg, G.L.~Kotkin, V.G.~Serbo, V.I.~Telnov,
{\it Nucl. Instr. \& Methods} {\bf 205} (1983) 47; {\it Sov. Yad. Fiz.}  {\bf 38} (1983) 372.


\bibitem{Telnov} V.I. Telnov, {\it Proc.  Workshop on Phys. and Experiments with Linear Colliders}, Saariselka,
Finland, 9-14/09, 1991, Ed. R.Orava et al.,World Sc. (1992), 739-776. Conf. Proc. {\bf  C9109093}
(1991) 739-775; {Proc. IX Int. Workshop  on Photon Photon Collisions}, San Diego,
22-26/03 1992, Ed. D.Caldwell, H.Paar, World Sc. (1992), 369-387.



\bibitem{Teln2015}  V.I. Telnov. Report at this session

\bibitem{Akerlof}
C.~Akerlof, {\it Preprint University of Michigan UM HE 81-59} (1991).

%\bibitem{KPS81}
%A.M.~Kondratenko, E.V.~Pakhtusova and E.L.~Saldin, Preprint INP
%81- 130 (1981), Novosibirsk; Dokl. Akad. Nauk 264 (1982) 3.


\bibitem{GKPS83gg}
I.F.~Ginzburg, G.L.~Kotkin, S.L.~Panfil and V.G.~Serbo, {\it Sov. Yad. Fiz.} {\bf 38} (1983) 614


\bibitem{GKPST84} I.~F.~Ginzburg,
G.~L.~Kotkin, S.~L.~Panfil, V.~G.~Serbo and V.~I.~Telnov,
{\it Nucl.\ Instrum.\ Meth.}\ {\bf A219} (1984) 5

\bibitem{TESLA} R.D. Heuer et al. {\em TESLA
Technical Design Report},  DESY 2001-011, TESLA
Report 2001-23, TESLA FEL 2001-05 (2001) p. 1--192
hep-ph/0106315; B. Badelek et al. The Photon
Collider at TESLA. {\it Int. J. Mod. Phys.} {\bf A 19}
(2004) 5097-5186


\bibitem{ILC}  International Linear Collider. Technical Design report (2007-2010-2013); M. Harrison, M. Ross, N. Walker, hep-ph/1308.3726


\bibitem{Ritus} A.I.~Nikishov, V.I.~Ritus, {\it Quantum Electrodynamics of Phenomena in
the Intense Electromagnetic Field} (Proc. of the Lebedev Institute,
in Russian) Moscow, {\bf 111} (1979)); A.I.~Nikishov, V.I.~Ritus, {\it Zh.
Eksp. Teor. Fiz.} {\bf 46} (1964) 776

\bibitem{GKPphot} I.F. Ginzburg, G.L. Kotkin, S.I. Polityko.
%On a possibility to observe and use production of high energy %photons by electrons in the field of intense electromagnetic wave.
{\it Sov. Yad. Fiz.\ }
{\bf 40} (1984) 1495-1499.

\bibitem{GKPel}  I.F. Ginzburg, G.L. Kotkin, S.I. Polityko.
%On possible observation of electron--positron pair production by a %photon in a field of intense electromagnetic wave.
{\it Sov. Yad. Fiz.\ } {\bf 37} (1983) 368.


\bibitem{IKP} D.Yu. Ivanov, G.L. Kotkin, V.G. Serbo. {\it Eur. Phys. J.} {\bf C36} (2004)127-145  arXiv:hep-ph/0402139;   0310325; 0311222;
    {\it Eur.Phys.J.} {\bf C40} (2005) 27-40



\bibitem{Telnovspectr} V. I. Telnov, {\it Nucl. Instrum. Meth.} {\bf A 355} (1995) 3;  {\it A Code for the simulation
of luminosities and QED backgrounds at photon colliders,} talk  at Second
Workshop of ECFA-DESY study, Saint Malo, France, April 2002.

\bibitem{ggamlum}  A.V. Pak, D.V. Pavluchenko, S.S. Petrosyan, V.G. Serbo, V.I. Telnov, {\it Nucl.Phys.} {\bf B (Proc.Suppl.) 126}:379-385,2004; arXiv:0301037[hep-ex]

\bibitem{GKrho} I.F.~Ginzburg, G.L.~Kotkin, {\it Eur. Phys. J.}  {\bf C 13} (2000) 295

\bibitem{Tel2001} V.I. Telnov, {\it  Nucl. Instrum. Meth.} {\bf A 472} (2001) 280-290.


\bibitem{GKPho9}  I.F.~Ginzburg, G.L.~Kotkin, Photon2009

\bibitem{GinKot15} I.F.~Ginzburg, G.L.~Kotkin, in preparation

\bibitem{Kotserpolar} G.L. Kotkin, H. Perlt, V.G. Serbo, {\it  Nucl. Instrum. Meth.} {\bf A404} (1998) 430-436;  arXiv:hep-ph/9706405


\bibitem{SerCher} V.L. Chernyak, V.G. Serbo, {\it Nucl.Phys.} {\bf B 67} (1973) 464;
I.F. Ginzburg, A.Schiller, V.G. Serbo,
{\it Eur. Phys. Journ.\ } {\bf C 18} (2001) 731-746

\bibitem{Witten}
E.~Witten, {\it Nucl. Phys.}, {\bf B120} (1977) 189.

\bibitem{GPS87}
I.F.~Ginzburg, S.L.~Panfil and V.G.~Serbo, {\it Nucl. Phys.} {\bf B 284}
(1987) 585;  {\bf B 296} (1988) 569; I.F.Ginzburg and D.Yu.~Ivanov, {\it Nucl. Phys.} {\bf B 388} (1992) 376;
{\it Nucl. Phys. B Proc. Suppl.} {\bf 25 B} (1992) 224.


\bibitem{WGKPS83}
I.F.~Ginzburg, G.L.~Kotkin, S.L.~Panfil and V.G.~Serbo, {\it Nucl.
Phys.} {\bf B 228} (1983) 285; (E), {\bf B 243} (1984) 550.
\bibitem{GinIl93}  I.F. Ginzburg, V.A.Ilyin, A.E.Pukhov, V.G.Serbo,
S.A.Shichanin. %The third order processes with W and Z production
%in $\gamma e$ and $\gamma \gamma$ collisions.
{\it Rus. Yad. Fiz.\ } {\bf 56} (1993) 57--63.

\bibitem{ZerGin} I.F. Ginzburg, I.P. Ivanov, {\it Eur. Phys. Journ.\ } {\bf C 22} (2001) 411-421; hep--ph/0004069.


\bibitem{Gin2015b} I.F. Ginzburg, arXiv: 1502.01797 [hep-ph]




\end{thebibliography}
\end{document}